\documentclass[aps,prd,twocolumn,10pt,superscriptaddress,nofootinbib,nobibnotes,longbibliography]{revtex4-1}

\usepackage{amssymb}
\usepackage{graphicx}
\usepackage{amsmath}
\usepackage{hyperref}
\usepackage{subfigure}
\usepackage{multirow}
\usepackage{setspace}
\usepackage{verbatim}
\usepackage{float}
\usepackage{color}
\usepackage{ulem}
\usepackage[utf8]{inputenc}
\usepackage[table,xcdraw]{xcolor}
\usepackage{makecell}
\usepackage{booktabs}

\begin{document}
	
	\title{Gravitational waves from vacuum bubbles: Ultraviolet dependence on wall thickness}
	
	\author{Jun-Chen Wang}
	\email{junchenwang@stu.pku.edu.cn}
	\affiliation{School of Physics, Peking University, Beijing 100871, China}
	
	\author{Shao-Jiang Wang}
	\email{schwang@itp.ac.cn}
	\affiliation{Institute of Theoretical Physics, Chinese Academy of Sciences, Beijing 100190, China}
	\affiliation{Asia Pacific Center for Theoretical Physics (APCTP), Pohang 37673, Korea}
	
	\author{Zi-Yan Yuwen}
	\email{yuwenziyan@itp.ac.cn}
	\affiliation{Institute of Theoretical Physics, Chinese Academy of Sciences, Beijing 100190, China}
	\affiliation{School of Physical Sciences, University of Chinese Academy of Sciences (UCAS), Beijing 100049, China}
	
	\begin{abstract}
		The gravitational wave (GW) spectrum from the first-order phase transition can be characterized by a few phenomenological parameters but with high degeneracies in model/data distinguishments. In this paper, we look into the high-frequency power law of the GW spectrum with preliminary numerical simulations for both quantum and semiclassical pictures of vacuum decay. We first reveal an anticorrelation of the high-frequency power law to a certain power of the ratio between the wall thickness and bubble radius at the onset of bubble collisions, which can be further approximated analytically by some other phenomenological model characteristics to break the model degeneracy.
	\end{abstract}
	\maketitle
	
	\section{Introduction}\label{sec:introduction}
	
	As the early Universe evolved, the spontaneous breaking of a continuous symmetry may have led to a cosmological first-order phase transition (FOPT)~\cite{Mazumdar:2018dfl,Hindmarsh:2020hop,Athron:2023xlk}. At the beginning of an FOPT, cosmic bubbles separating false and true vacua can nucleate stochastically via vacuum decay in field theory. In addition to the well-known channel via quantum tunneling~\cite{Steinhardt:1981ct,Coleman:1977py,Callan:1977pt}, a semiclassical one is also plausible to realize vacuum decay~\cite{Braden:2018tky}. The former is purely a quantum effect allowing for barrier penetration with lower energy, while the latter could originate from the classical evolution of occasionally developed perturbations~\cite{Hertzberg:2019wgx,Blanco-Pillado:2019xny,Wang:2019hjx,Huang:2020bzb}. After nucleations, bubbles expand rapidly~\cite{Cai:2020djd,Wang:2022txy,Wang:2023kux,Yuwen:2024hme} and finally percolate via violent collisions, during which stochastic gravitational wave backgrounds (SGWBs) are generated~\cite{Caprini:2015zlo,Caprini:2019egz} via bubble collisions~\cite{Witten:1984rs,Hogan:1986qda}, sound waves~\cite{Hogan:1986qda}, and magneto-hydrodynamic (MHD) turbulences~\cite{Witten:1984rs} to be detectable by currently running GW detections (LIGO-Virgo~\cite{Romero:2021kby,Huang:2021rrk,Jiang:2022mzt,Badger:2022nwo,Yu:2022xdw}, PTAs~\cite{Xu:2023wog,NANOGrav:2023gor,EPTA:2023sfo,Reardon:2023gzh}) and future planning GW detectors (LISA~\cite{LISA:2017pwj,Baker:2019nia}, Taiji~\cite{Hu:2017mde,Ruan:2018tsw}, TianQin~\cite{TianQin:2020hid,TianQin:2015yph}).
	
	Detecting and analyzing the energy spectrum of this SGWB offers a promising way to probe the FOPT and its underlying new physics~\cite{Cai:2017cbj,Bian:2021ini}, which requires accurately modeling the GW power spectrum with analytical templates as informative as possible. The GW spectrum can be speculatively fitted with a few phenomenological parameters from numerical simulations of bubble wall collisions~\cite{Kosowsky:1991ua,Kosowsky:1992rz,Kosowsky:1992vn,Kamionkowski:1993fg,Huber:2008hg,Konstandin:2017sat,Cutting:2018tjt,Cutting:2020nla}, sound waves~\cite{Hindmarsh:2013xza,Hindmarsh:2015qta,Hindmarsh:2017gnf,Cutting:2019zws,Sharma:2023mao}, and MHD turbulences~\cite{Kamionkowski:1993fg,Niksa:2018ofa,Pol:2019yex,Brandenburg:2021tmp,Brandenburg:2021bvg,Auclair:2022jod,Dahl:2024eup} but under the help of analytical understandings of thin-wall collisions~\cite{Caprini:2007xq,Caprini:2009fx,Jinno:2016vai,Jinno:2017fby,Konstandin:2017sat}, sound shell collisions~\cite{Hindmarsh:2016lnk,Hindmarsh:2019phv,Guo:2020grp,Cai:2023guc,RoperPol:2023dzg}, and vortical turbulences~\cite{Kosowsky:2001xp,Dolgov:2002ra,Nicolis:2003tg,Caprini:2006jb,Gogoberidze:2007an,Caprini:2009yp}, respectively. In particular, unlike the strongly coupled case with a nonrelativistic terminal wall velocity reached long before collisions~\cite{Li:2023xto,Wang:2023lam}, the GW contribution from wall collisions could dominate over that of sound waves when most bubbles collide long before they would have approached the nearly constant terminal velocity~\cite{Cai:2020djd,Lewicki:2022pdb}.
	
	However, the high-frequency tail in the GW power spectrum has not been fully understood yet. Taking the simplest case of vacuum phase transitions without thermal fluids as an example, previous analytical modelings of thin-wall collisions with~\cite{Jinno:2016vai} and without~\cite{Jinno:2017fby} the envelope approximation agree on an $k^{-1}$ power law for the high-frequency part of the GW spectrum, but a semianalytical simulation from a dubbed bulk flow model~\cite{Konstandin:2017sat} also beyond the envelope approximation produces rather different $k^{-2}$ and $k^{-3}$ power laws at ultraviolet (UV) for ultrarelativistic and nonrelativistic walls, respectively. Further relaxing the thin-wall approximation, a previous numerical simulation~\cite{Cutting:2020nla,Gould:2021dpm} has found a steeper UV power than $k^{-1}$ depending roughly on a model parameter that is intimately related to the bubble wall thickness. 
	
	In this paper, we will make this relation clearer by preliminarily quantifying this relation between the high-frequency power law of GW spectrum and bubble wall thickness at collision.  Following Refs.~\cite{Braden:2014cra,Gould:2021dpm,Jinno:2019bxw,Lewicki:2019gmv,Cutting:2020nla}, we first numerically simulate the bubble nucleations, expansion, and collisions in Sec.~\ref{sec:nucleation} and Sec.~\ref{sec:collision}, and then calculate and extract the GW spectrum in Sec.~\ref{sec:gravitational}. In doing so, we analytically derive in Sec.~\ref{sec:distance} for the first time the mean bubble separation at nucleation for runaway bubbles from vacuum phase transitions, where bubbles continue to accelerate without approaching a constant terminal velocity at collisions. Regardless of the nucleation channels of quantum or semiclassical types, we find in Sec.~\ref{sec:fitting} the high-frequency power law decreases to a certain power of the wall thickness at collisions. Finally in Sec.~\ref{sec:expression}, based on the total energy conservation, we analytically approximate the wall thickness at collisions as a function of other more model-independent parameters, such as the phase hierarchy $\Delta\phi/M_\mathrm{Pl}$, strength factor $\alpha=\Delta V/\rho_\mathrm{rad}$, inverse duration $\beta/H$, and wall velocity $v_w$ as we will define later. This informative modeling of the high-frequency power law together with the remaining parts of the GW spectrum largely breaks model degeneracies for future GW constraints on FOPTs. The last section~\ref{sec:conclusion} is devoted to conclusions and discussions.

	\section{Bubble nucleation}\label{sec:nucleation}
	
\begin{figure}
		\includegraphics[width=0.49\textwidth]{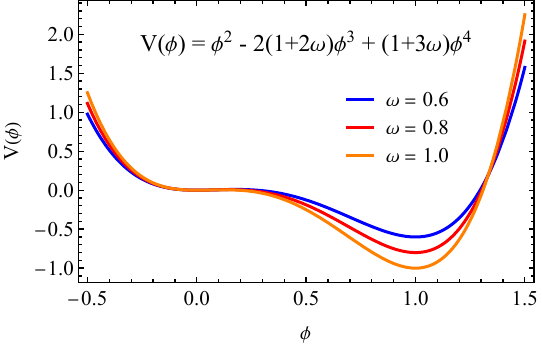}
		\caption{The scalar potential \eqref{eq:scalar_potential} used in the numerical simulations for various values of the parameter $\omega$.}
		\label{fig:potential}
	\end{figure}

	For simplicity, we follow Ref.~\cite{Wang:2019hjx} to consider a polynomial scalar potential with two minima as shown in Fig.~\ref{fig:potential},
	\begin{equation}
		\label{eq:scalar_potential}
		V(\phi;\omega>0)=\phi^2 - 2(1+2\omega)\phi^3 + (1+3\omega)\phi^4.
	\end{equation}
	The global minimum is at $\phi_{-}\equiv 1$ with $V(\phi_-)\equiv V_- =-\omega$, while the local minimum is at $\phi_+\equiv 0$ with $V(\phi_+)\equiv V_{+}=0$. Note that all dimensional quantities have been normalized with $\phi_-$. The potential difference between the false and true vacua is given by $\Delta V \equiv V_+ - V_- = \omega$. The height of the potential barrier, $V_b \equiv V_m - V_+$, is the difference between $V_+$ and the local maximum of the potential barrier, $V_m = \frac{1}{16}(1+4\omega)/(1+3\omega)^3$ at $\phi_m = 1/(2+6\omega)$. Typically, the initial thickness of bubble wall can be evaluated as $\Delta V/V_b$~\cite{Blanco-Pillado:2019xny}, which, in our case~\eqref{eq:scalar_potential}, reads $\Delta V/V_b=16\omega(1+3\omega)^3/(1+4\omega)$, rendering the bubble wall from a thin wall to a thick wall as $\omega$ increases.

\begin{figure*}
		\includegraphics[width=0.49\textwidth]{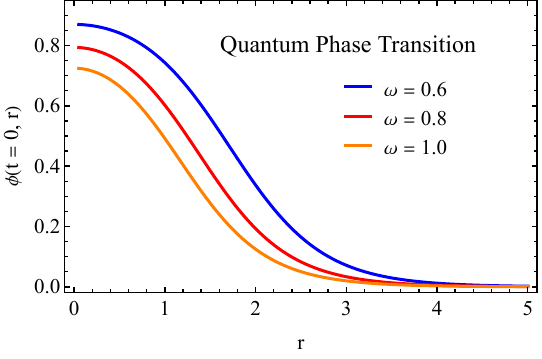}
		\includegraphics[width=0.49\textwidth]{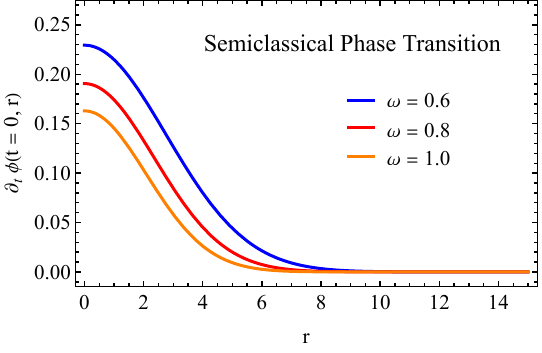}
		\caption{The initial scalar profiles for the quantum phase transition (left) and semiclassical phase transition (right) with $A/A_0=1$ and $R/R_0=1$ for a typical illustration.}
		\label{fig:initial_profile}
	\end{figure*}

	Once the potential is determined, the scalar field profile at the time of bubble nucleation can be obtained, serving as the initial condition for the subsequent numerical simulations. Typically, the bubble nucleation is understood as a quantum tunneling process, and the tunneling rate $\Gamma$ can be computed using the Euclidean instanton approach~\cite{Coleman:1977py,Callan:1977pt,Linde:1981zj}, with $\Gamma \propto e^{-S_E}$, where $S_E$ is the Euclidean action of the scalar field,
	\begin{equation}
		\label{eq:Euclidean_action}
		S_E = \int \mathrm{d}^4 x \left[\frac{1}{2}\left(\partial_\tau\phi\right)^2+\frac{1}{2}\left(\nabla \phi\right)^2+V(\phi)\right].
	\end{equation}
	The scalar profile $\phi(r=\sqrt{\tau^2+\mathbf{x}^2})$ that minimizes the action~\eqref{eq:Euclidean_action} has the maximum tunneling probability, which can be solved numerically via a public \textit{Mathematica} code \texttt{AnyBubble}~\cite{Masoumi:2016wot} from the equation of motion (EoM)
	\begin{align}\label{eq:Euclidean_equation}
		\nabla^2\phi=\frac{\mathrm{d}^2\phi}{\mathrm{d}r^2}+\frac{3}{r}\frac{\mathrm{d}\phi}{\mathrm{d}r}=\frac{\mathrm{d}V}{\mathrm{d}\phi}
	\end{align}
	under boundary conditions $\phi'(r=0) =0$ and $\phi(r \to \infty) \to \phi_+\equiv 0$. Thus, the solution to EoM~\eqref{eq:Euclidean_equation} exhibits $\mathcal{O}(4)$ symmetry and the corresponding profiles are displayed in the left panel of Fig.~\ref{fig:initial_profile} for different $\omega$ values. We can find that $\phi(\tau=0) \sim \phi_-$ near $\mathbf{x}=0$ and $\phi(\tau=0)$ tends to $\phi_+$ as $\mathbf{x}$ increases. Additionally, $\phi(\tau=0,\mathbf{x}=0)$ is closer to $\phi_-$ for a smaller $\omega$, that is, the thin-wall limit.

	In addition to quantum tunneling, bubble nucleation can also occur through a semiclassical mechanism~\cite{Braden:2018tky,Hertzberg:2019wgx,Blanco-Pillado:2019xny,Wang:2019hjx,Huang:2020bzb}. In this scenario, quantum fluctuations \cite{Hertzberg:2019wgx}, thermal fluctuations \cite{Berera:1995wh} or gravitational effects \cite{Blanco-Pillado:2019xny} drive the scalar field or its time derivative to develop a nonzero profile. The classical evolution of these fluctuations then leads to the bubble formation in real space instead of the Euclidean space~\cite{Braden:2018tky}. One of the simplest realizations~\cite{Blanco-Pillado:2019xny,Wang:2019hjx} of this semiclassical mechanism is the time derivative of the scalar field occasionally develops a Gaussian profile large enough to eventually overcome the potential barrier directly from a homogeneous field at the false vacuum. The initial condition for bubble formation is given by
	\begin{equation}
		\label{eq:initial_semiclassical}
		\phi(t=0,r)=0,\ \ \ \partial_t \phi(t=0,r)=A\exp\left(-\frac{r^2}{2R^2}\right),
	\end{equation}
	where $A$ and $R$ represent the amplitude and width of the Gaussian fluctuations, respectively. For a bubble to actually expand, the condition $R>R_0$ should at least be required, where $R_0 \equiv 2\sigma / |\Delta V|$ is the minimal expansion radius with the bubble wall tension $\sigma$ defined by
	\begin{equation}
		\label{eq:wall_tension}
		\sigma = \int_{\phi_+}^{\phi_-}\sqrt{2(V(\phi)-V_-)}\mathrm{d}\phi.
	\end{equation}
	To uniformly specify the Gaussian profile in \eqref{eq:initial_semiclassical}, we normalize it as
	\begin{equation}
		\label{eq:initial_semiclassical_new}
		\partial_t \phi(t=0,r)=\left(\frac{A}{A_0}\right)A_0 \exp\left[-\frac{(r/R_0)^2}{2(R/R_0)^2}\right],
	\end{equation}
	where $A_0=\sqrt{2eV_b}$ is defined in such a way that the critical velocity $\dot{\phi}_c\equiv \sqrt{2V_b}$ is realized exactly at $r=R$ for $A=A_0$. The profiles of \eqref{eq:initial_semiclassical_new} are shown in the right panel of Fig.~\ref{fig:initial_profile}. Ref.~\cite{Wang:2019hjx} provides a detailed discussion on the ranges of $A/A_0$ and $R/R_0$ that permit bubble nucleation. In our subsequent calculations, we exactly follow the same configuration and procedure as Ref.~\cite{Wang:2019hjx} to generate the semiclassical bubble from simply fixing $A/A_0 = 1$ and $R/R_0 = 1$ within the range where bubble nucleation can occur. The specific choice of parameters $A$ and $R$ within this allowable range does not affect our final results.

	Having established all initial scalar profiles, we can proceed with numerical simulation to evolve the bubble expansion until the collision. In the following, we will refer to scenarios where bubbles nucleate via quantum tunneling as ``quantum phase transitions'' and those nucleating through the semiclassical mechanism as ``semiclassical phase transitions.''
	
	\section{Bubble separation}\label{sec:distance}
	After obtaining the scalar field initial profiles from quantum tunneling and semiclassical mechanisms, there is one last quantity to be determined for the initial configuration of numerical simulation: the mean bubble separation at the nucleation time, which should be twice the mean bubble radius at collisions. In this study, we focus on the vacuum FOPT, and hence, the potential~\eqref{eq:scalar_potential} is independent of the background temperature. In such a scenario, the bubble nucleation rate $\Gamma$ is time-independent and given by~\cite{Coleman:1977py,Callan:1977pt,Linde:1980tt,Linde:1981zj,Cai:2017tmh}\ \footnote{It is important to emphasize that \eqref{eq:nucleation_rate} represents only an approximate expression for the bubble nucleation rate, neglecting functional determinant factors and higher-order loop corrections. Specifically, this approximation becomes invalid in the thin-wall limit, for which the correct form of the nucleation rate can be found, for example, in Ref.~\cite{Garriga:1994ut}. However, since our numerical calculations do not explore the thin-wall limit, the approximation~\eqref{eq:nucleation_rate} provides a sufficiently accurate description for the scenarios considered in this paper.}
	\begin{equation}
		\label{eq:nucleation_rate}
		\Gamma \approx \frac{1}{R_0^4}\left(\frac{S_E}{2\pi}\right)^2 e^{-S_E},
	\end{equation}
	where $S_E$ and $R_0$ are the Euclidean action and initial bubble radius, respectively. Following Ref.~\cite{Cutting:2020nla}, we define the later bubble radius $R(t)$ by $\phi(t,R(t))=\phi_0/2$, and define the ``wall'' of the bubble as the region between $r_{\rm in}(t)$ and $r_{\rm out}(t)$, where $\phi(t,r_{\rm in}(t))=\phi_{0}(1-\tanh(-1/2))/2$ and $\phi(t,r_{\rm out}(t))=\phi_0(1-\tanh(1/2))/2$. Here, $\phi_0$ represents the scalar field value at the bubble center. Then, the wall thickness is defined as $l(t)\equiv r_{\rm out}(t)-r_{\rm in}(t)$.
	
	In the vacuum FOPT, the bubble expands without bound and runs away to approach the speed of light, and hence, there is no constant terminal wall velocity $v_w$ by the time of bubble collisions. In this case, the false vacuum fraction $F(t)$ is expressed as~\cite{Guth:1982pn,Turner:1992tz}
	\begin{equation}
		\label{eq:fraction_3}
		F(t)=e^{-\int_{t_i}^{t}\mathrm{d}t'\Gamma(t')V_{\rm b}(t,t')},
	\end{equation}
	where $t_i$ is the nucleation time of the first bubble and $V_{\rm b}(t,t')$ denotes the volume at time $t$ of a bubble nucleated at time $t'$,
	\begin{equation}
		\label{eq:volume}
		V_{\rm b}(t,t')=\frac{4}{3}\pi \left(\int_{t'}^{t}v_{w}(\tilde{t})\mathrm{d}\tilde{t}\right)^3.
	\end{equation}
	For a vacuum FOPT, we have (see, for example, Ref.~\cite{Cai:2020djd})
	\begin{equation}
		\label{eq:velocity_1}
		v_w(t)=\sqrt{1-\frac{R_0^2}{R(t)^2}},
	\end{equation}
	where $R(t)$ is the bubble radius at time $t$.
	Using $v_w(t)=\mathrm{d}R/\mathrm{d}t$, we can obtain
	\begin{align}
		\label{eq:radius_1}
		R(t)&=\sqrt{(t-t')^2+R_0^2},\\
		\label{eq:velocity_2}
		v_w(t)&=\frac{t-t'}{\sqrt{(t-t')^2+R_0^2}}.
	\end{align}
	Substituting \eqref{eq:velocity_2} into \eqref{eq:volume}, we get
	\begin{align}
		\label{eq:volume_1}
		V_{\rm b}(t,t')&=\frac{4}{3}\pi \left(\int_{t'}^{t}\frac{\tilde{t}-t'}{\sqrt{(\tilde{t}-t')^2+R_0^2}}\mathrm{d}\tilde{t}\right)^3\\ \notag
		&=\frac{4}{3}\pi \left(\sqrt{(t-t')^2+R_0^2}-R_0\right)^3.
	\end{align}
	Thus, the false vacuum fraction $F(t)$ becomes
	\begin{align}
		\label{eq:fraction_4}
		F(t)&=e^{-\frac{4}{3}\pi \Gamma \int_{t_i}^{t}\mathrm{d}t'\left(\sqrt{(t-t')^2+R_0^2}-R_0\right)^3}\nonumber\\
		&=e^{-\frac{\pi}{6} \Gamma R_0^4 \mathcal{A}\left(\frac{t-t_i}{R_0}\right)}, \\ 
		\label{eq:A_function}
		\mathcal{A}(x)&=-32x-8x^3+17x\sqrt{1+x^2}+2x^3\sqrt{1+x^2}\nonumber\\
		&+15\ln \left(x+\sqrt{1+x^2}\right).
	\end{align}
	Now the bubble number density $n_{\rm b}$ can be determined,
	\begin{align}
		\label{eq:number_2}
		n_{\rm b}=&\int_{t_i}^{t_f} \Gamma(t)F(t) \mathrm{d}t=\Gamma \int_{t_i}^{t_f} e^{-\frac{\pi}{6}\Gamma R_0^4\mathcal{A}\left(\frac{t-t_i}{R_0}\right)}\mathrm{d}t \\ \notag
		\simeq & \Gamma \int_{t_i}^{\infty} e^{-\frac{\pi}{6}\Gamma R_0^4\mathcal{A}\left(\frac{t-t_i}{R_0}\right)}\mathrm{d}t
		=\Gamma R_0 \int_{0}^{\infty} e^{-\frac{\pi}{6}\Gamma R_0^4\mathcal{A}(x)}\mathrm{d}x.
	\end{align}
	Here, the reference time $t_f$ is defined by $F(t_f)\equiv1/e$ and we approximate $t_f$ as infinity when using saddle-point approximation in~\eqref{eq:number_2}. Finally, the mean bubble separation $d$ is obtained as
	\begin{equation}
		\label{eq:distance_value}
		d=2R_{\rm col} =\frac{2}{n_{\rm b}^{1/3}},
	\end{equation}
	where $R_{\rm col}$ is the mean bubble radius at the collision time. For a vacuum FOPT, once the values of $R_0$ and $S_E$ are known, the mean bubble distance at the nucleation time can be derived by~\eqref{eq:nucleation_rate}, \eqref{eq:number_2}, and~\eqref{eq:distance_value}. The integral in \eqref{eq:number_2} should be calculated numerically. However, when $\pi \Gamma R_0^4/6 \gg 1$, which corresponds to a thick wall bubble, the integration can be approximated analytically using the saddle-point method. Specifically, we can expand the function $A(x)$ at $x=0$ as
	\begin{equation}
		\label{eq:At_function_expand}
		A(x)=\frac{1}{7}x^7+\mathcal{O}(x^9).
	\end{equation}
	Substituting this approximation into \eqref{eq:number_2}, we obtain
	\begin{align}
		\label{eq:number_4}
		n_{\rm b}\simeq& \Gamma R_0 \int_0^{\infty}e^{-\frac{\pi}{42}\Gamma R_0^4 x^7}\mathrm{d}x \\ \notag
		=&\Gamma_{\rm E}\left(\frac{8}{7}\right)\left(\frac{42}{\pi}\right)^{1/7}\Gamma^{6/7}R_0^{3/7},
	\end{align}
	where $\Gamma_{\rm E}$ is the Euler-Gamma function and $\Gamma_{\rm E}(8/7)\simeq 0.935$. In Table~\ref{tab:distance}, we present the mean bubble separation for various values of the potential parameter $\omega$. It is observed that, when $\omega < 0.6$, the bubble separation becomes extremely large, and simulating such a system would require significant computational resources.  Therefore, in this study, we restrict our simulations to values of $\omega$ ranging from 0.6 to 1.0, with an interval of 0.05.
	
	\begin{table}[ht]
		\centering
		\renewcommand{\arraystretch}{1.35} 
		\setlength{\tabcolsep}{4pt} 
		\textbf{Mean bubble separation} \\[0.5em]
		\scalebox{1.2}{
			\begin{tabular}{*{12}{c}}
				\hline
				\hline
				$\omega$  & 0.10 & 0.20 & 0.30 & 0.40 &0.50 & 0.60 & 0.65 \\
				\hline
				$d$ & $10^{76}$ & $10^{12}$  & $10^{5}$ & 776 & 88.6 & 29.7 & 21.0 \\
				\hline
				\hline
				$\omega$  & 0.70 & 0.75 & 0.80 & 0.85 &0.90 & 0.95 & 1.00 \\
				\hline
				$d$ & 16.0 & 13.1 & 11.0 & 9.72 &  8.82 & 8.13 & 7.61 \\
				\hline
				\hline
		\end{tabular}}
		\caption{The mean bubble distance $d$ at the nucleation time in the quantum phase transition for different $\omega$ values.}
		\label{tab:distance}
	\end{table}
	
	Lastly, our analysis above focuses on the mean bubble separation in the quantum phase transition, which is determined by the bubble nucleation rate. For the semiclassical phase transition, the bubble nucleation rate and the corresponding mean separation have been extensively studied in Refs.\cite{Braden:2018tky,Hertzberg:2020tqa,Tranberg:2022noe}. In particular, Ref.~\cite{Braden:2018tky} has corrected and updated their results in an erratum with a factor of 2 difference in their plotting code as pointed out by Ref.~\cite{Hertzberg:2020tqa}, which causes an enhancement in the nucleation rate by a factor of $1/4$ in the slope of $\ln\Gamma$. In this work, when simulating the semiclassical phase transition, we will simply use the same bubble separation values presented in Table~\ref{tab:distance} but with such a difference in mind that we are not comparing the quantum and semiclassical phase transitions on the same footing.

	\section{Bubble expansion and collision}\label{sec:collision}
	
	After deriving the scalar initial profiles and the bubble distance, we closely follow Ref.~\cite{Lewicki:2019gmv} to numerically simulate two-bubble expansion and collision as the simplest representative for later extraction of GW signals. The EoM for the scalar field is still of the form
	\begin{equation}
		\label{eq:EoM_origin}
		\nabla^2\phi=\frac{\mathrm{d}V(\phi)}{\mathrm{d}\phi}
	\end{equation}
	but now both evolve in the real space with initial profiles containing two bubbles. For simplicity, we assume the two bubbles are identical. When the bubbles are separated by a distance $d$, their centers can always be positioned on the $z$-axis at $z = \pm d/2$ by adjusting the coordinate system. Now, the evolution of this two-bubble system is best described in cylindrical coordinates $(t, \rho, \theta, z)$, in which case the Eq.~\eqref{eq:EoM_origin} is expanded as
	\begin{equation}
		\label{eq:EoM}
		\partial_t^2\phi-\partial_{\rho}^2\phi-\frac{1}{\rho}\phi-\partial_z^2\phi=-\frac{\mathrm{d}V(\phi)}{\mathrm{d}\phi}.
	\end{equation}
	During our simulation, we discretize the space using a lattice with 1600 grid points per spatial dimension. The entire range for $z$ and $\rho$ is set between $-1.5d$ and $1.5d$, and from 0 to $1.5d$, respectively. For the time dimension, we divide it into $10^5$ slices and simulate the bubble evolution using the sixth-order Runge-Kutta method. In our simulations, the total time extent $t_{\rm ext}$ is determined by the bubble separation $d$. Specifically, we set $t_{\rm ext} \simeq d$ for the quantum phase transition scenario, and $t_{\rm ext} \simeq 1.5d$ for the semiclassical case, accounting for the additional time required for bubble nucleation. Note that this choice of $t_{\rm ext} = \mathcal{O}(d)$ is further motivated by the physical situation, where more bubbles coming into contact with the two-bubble system. It is important to emphasize that, as thoroughly demonstrated in Ref.~\cite{Kosowsky:1991ua}, the GW spectrum is strongly dependent on $t_{\rm ext}$, particularly its peak amplitude. Nevertheless, consistent with findings in Ref.~\cite{Lewicki:2019gmv}, certain special properties, such as the high-frequency power we care about in this study, exhibit relatively weak sensitivity to the precise value of $t_{\rm ext}$.

\begin{figure}
		\includegraphics[width=0.45\textwidth]{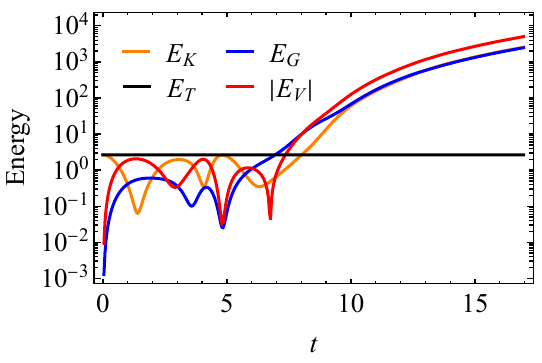}
		\caption{The time evolution of the kinetic energy $E_K$, gradient energy $E_G$, potential energy $E_V$ (in absolute value), and total energy $E_T$ for one of the numerical simulation results of bubbles nucleated via the semiclassical mechanism. The potential parameter is $\omega=0.8$, and the initial condition parameters are $A/A_0=1$ and $R/R_0=1$.}
		\label{fig:energy}
	\end{figure}

	For the consistency check of the numerical stability, we have calculated the full time evolution of the total kinetic energy $E_K$, gradient energy $E_G$, and potential energy $E_V$ during the whole simulation,
	\begin{align}
		\label{eq:kinetic_energy}
		E_{K}&=\int\mathrm{d}\mathcal{V}\frac{(\partial_t \phi)^2}{2}=2\pi \int_{-\infty}^{\infty}\mathrm{d}z\int_{0}^{\infty}\mathrm{d}\rho\ \frac{\rho}{2}(\partial_t\phi)^2,\\
		\label{eq:gradient_energy}
		E_{G}&=\int\mathrm{d}\mathcal{V}\frac{(\nabla \phi)^2}{2} \\ \nonumber
		&=2\pi \int_{-\infty}^{\infty}\mathrm{d}z\int_0^{\infty}\mathrm{d}\rho\ \frac{\rho}{2}[(\partial_z \phi)^2+(\partial_{\rho} \phi)^2],\\
		\label{eq:potential_energy}
		E_{V}&=\int \mathrm{d}\mathcal{V} V(\phi) =2\pi \int_{-\infty}^{\infty}\mathrm{d}z\int_0^{\infty}\mathrm{d}\rho\ \rho V(\phi),
	\end{align}
	where $\mathrm{d}\mathcal{V}$ is the volume element integrated over the open regions of true bubbles. Throughout the numerical simulations, the total energy $E_T \equiv E_K + E_G + E_V$ should remain conserved (or vanished specifically for the quantum phase transition). One of our numerical results for the time evolution of semiclassical bubble expansion is shown in Fig.~\ref{fig:energy}, where it can be seen that $E_K$, $E_G$, and the absolute value of $E_V$ are converged into a constantly evolving total energy $E_T$. Specifically, in our simulations, the relative deviation between the total energy $E_T$ at the initial and final times is only at the order of $\mathcal{O}(10^{-4})$. For quantum phase transitions, the numerical stability has also been checked with $E_T=0$ since the converted potential energy from the false vacuum to true vacuum has all been deposited into the kinetic and gradient energies of the bubble wall.
	
	\begin{figure*}
		\includegraphics[width=0.49\textwidth]{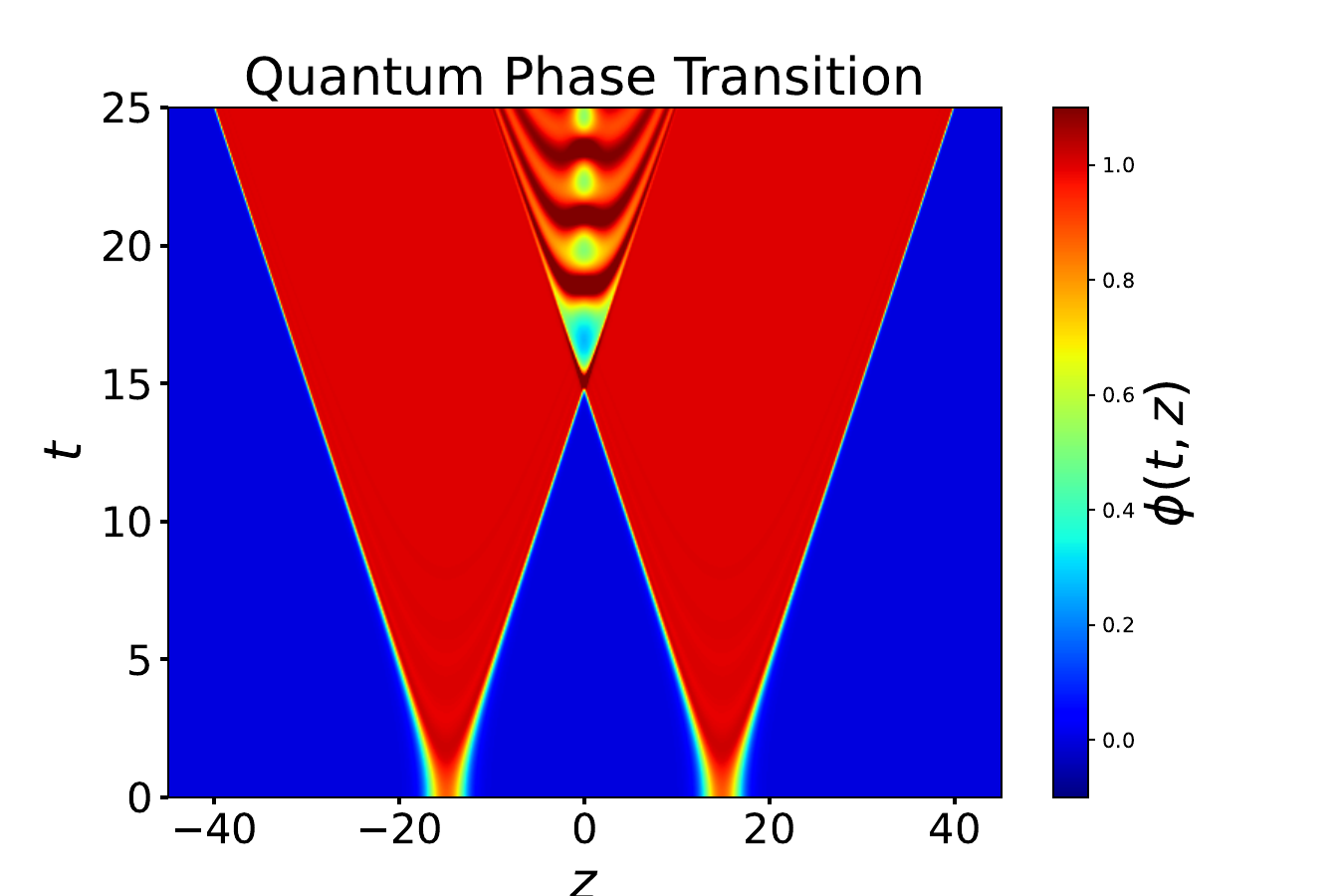}
		\includegraphics[width=0.49\textwidth]{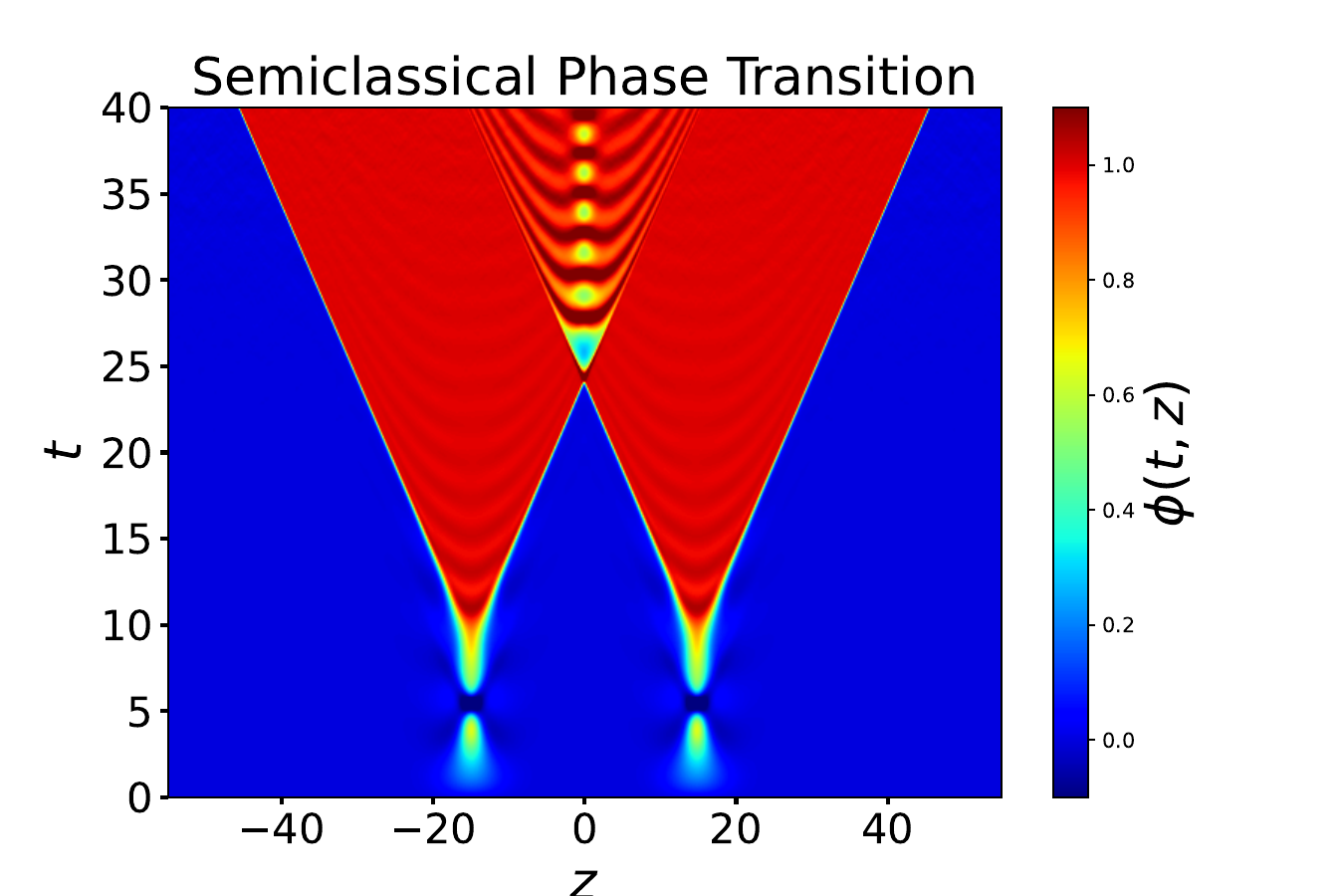}
		\includegraphics[width=0.49\textwidth]{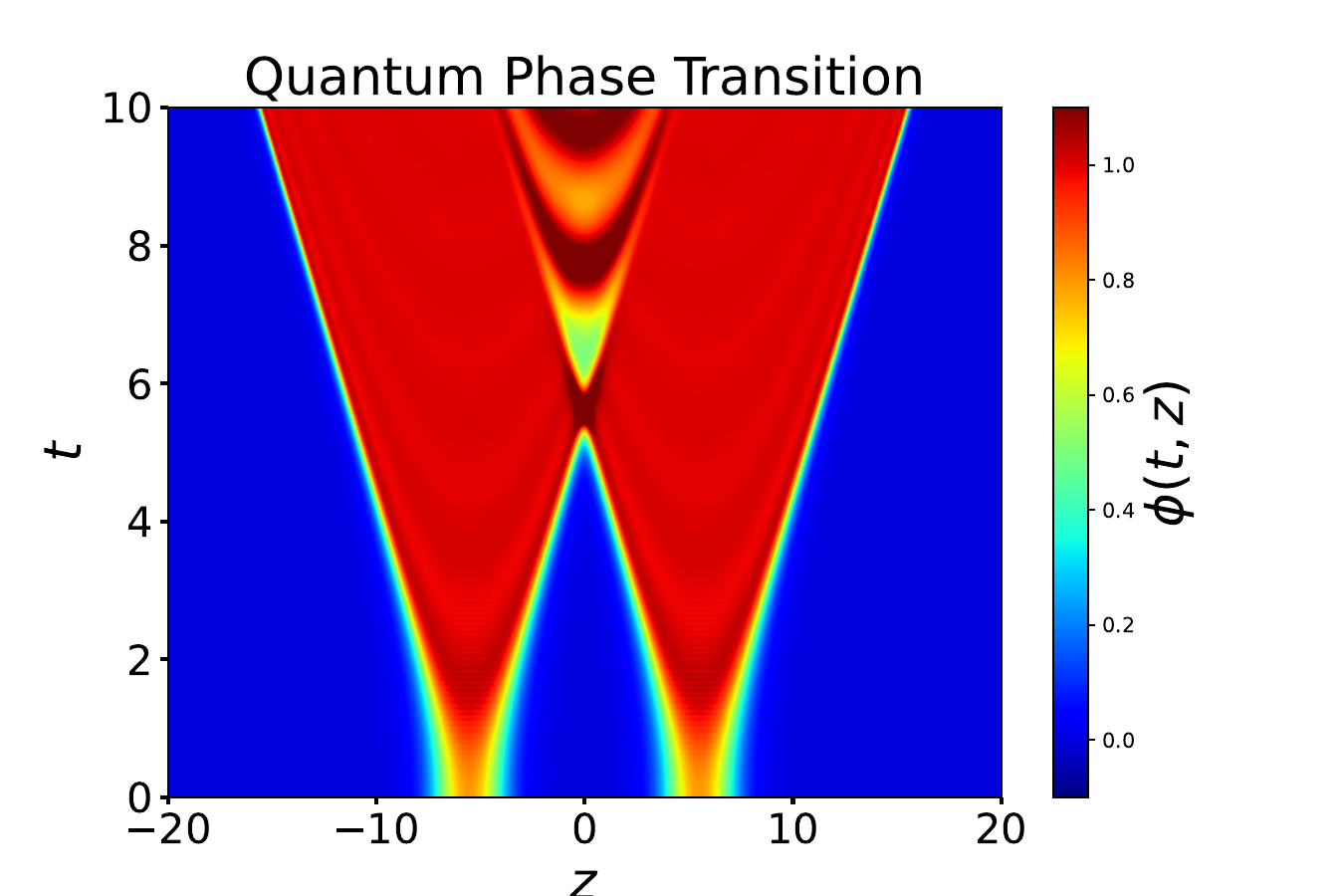}
		\includegraphics[width=0.49\textwidth]{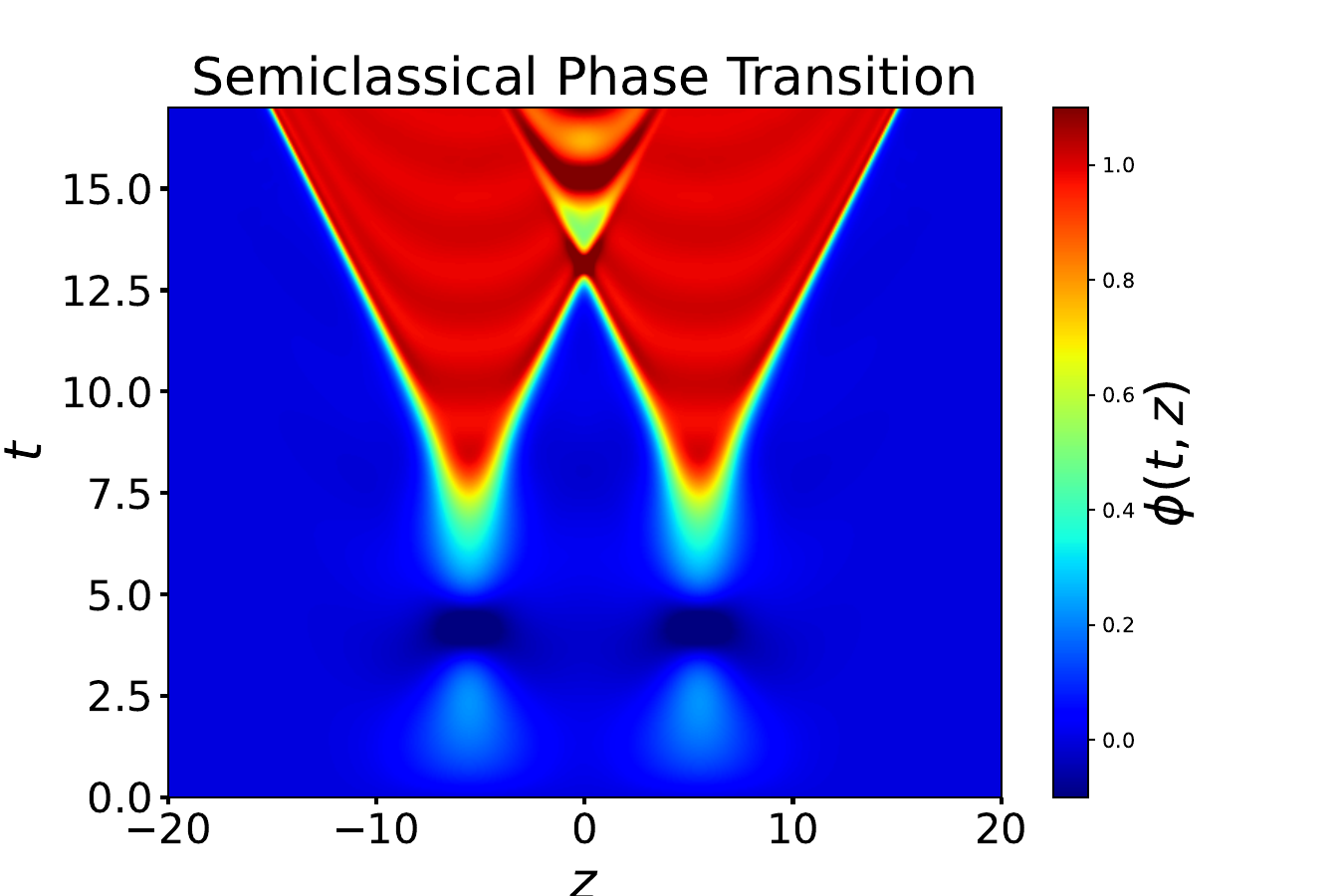}
		\includegraphics[width=0.49\textwidth]{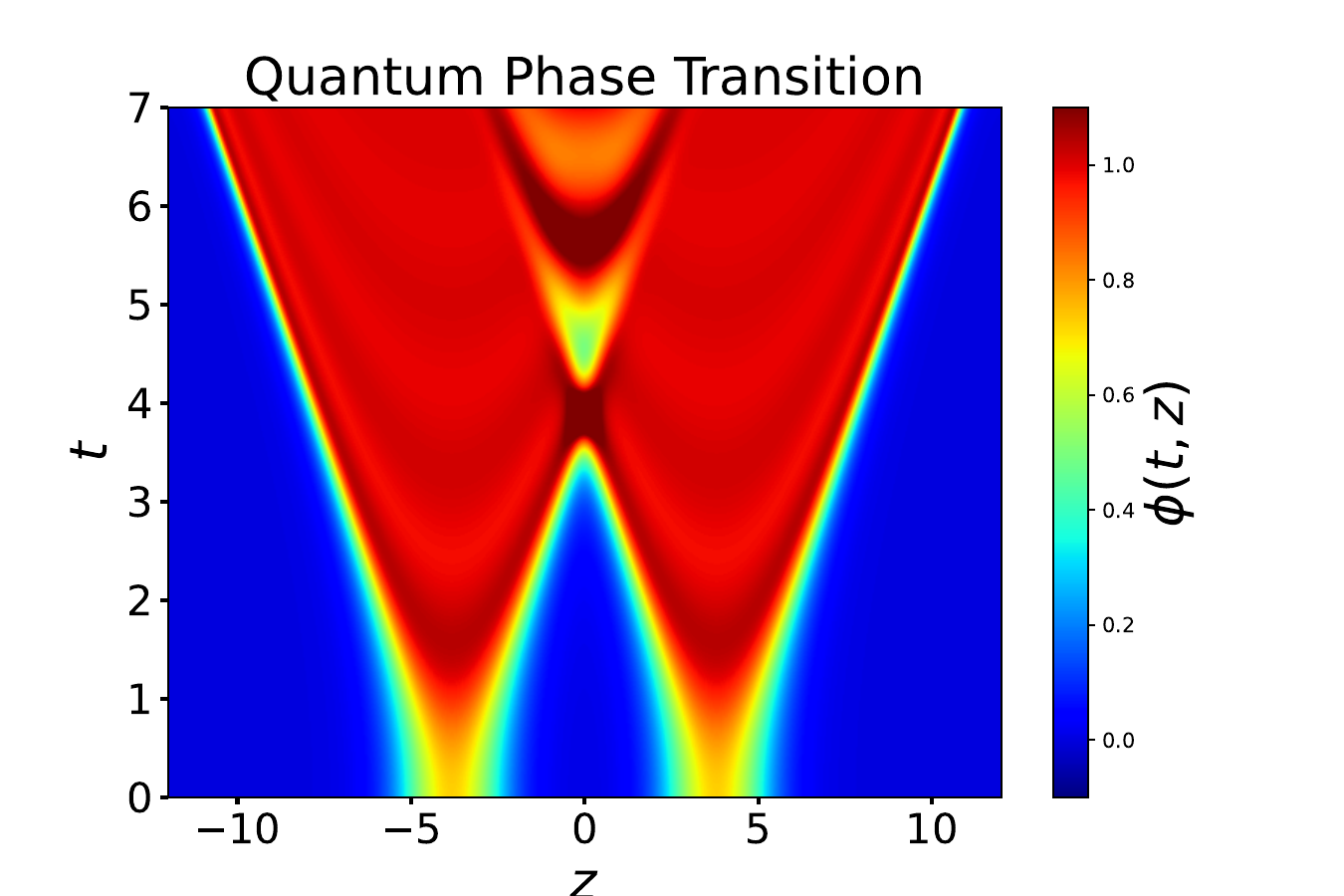}
		\includegraphics[width=0.49\textwidth]{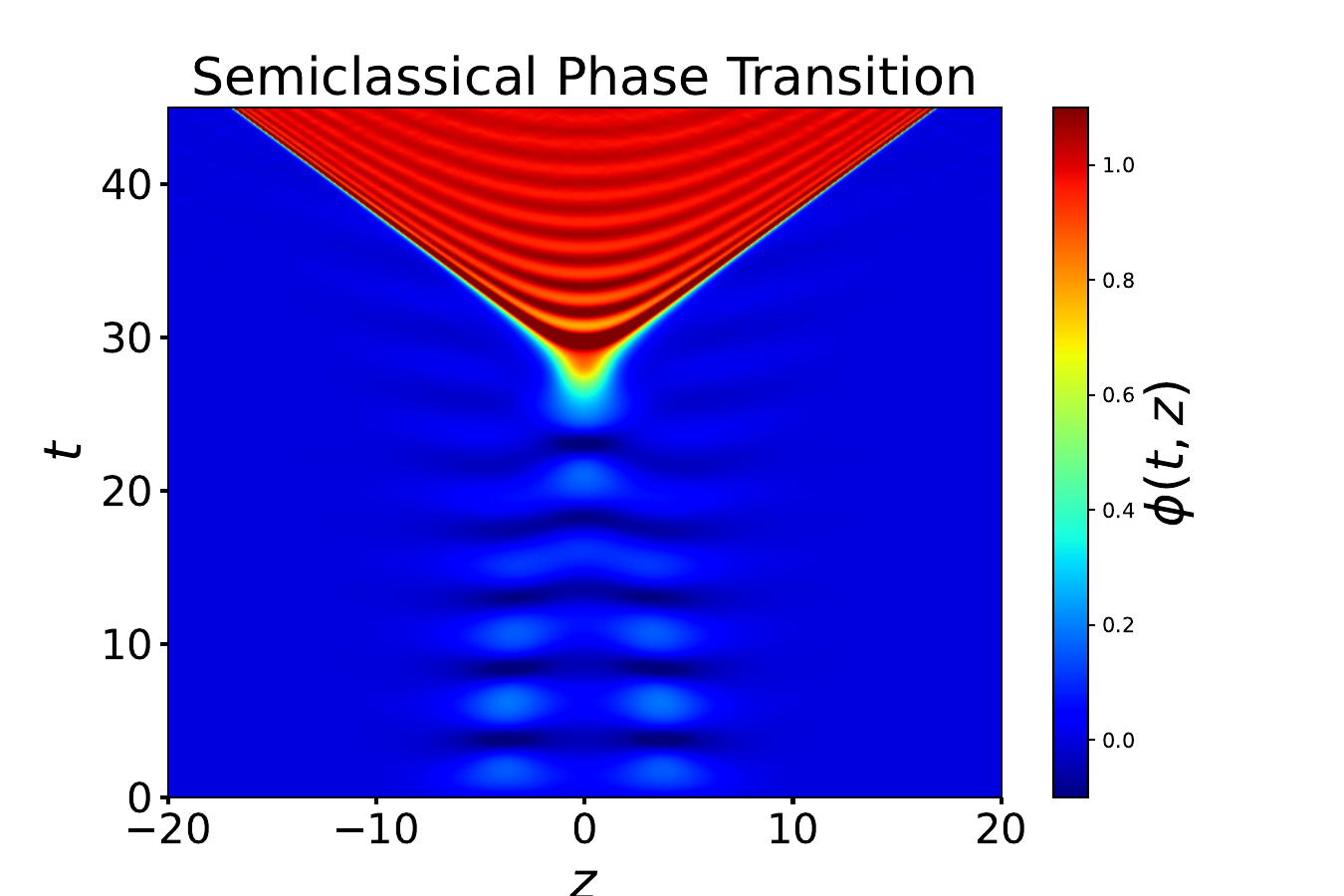}
		\caption{The simulation results of the bubble expansion and collision at $\rho=0$. The left and right panels correspond to bubbles originating from the semiclassical mechanism and quantum tunneling, respectively. The top, middle, and bottom panels correspond to the potential parameters $\omega = 0.6$ (thin wall), $\omega = 0.8$ (middle wall), and $\omega=1.0$ (thick wall), respectively.}
		\label{fig:bubble_profile}
	\end{figure*}

	We now compare the simulation results of two-bubble expansion and collision in quantum and semiclassical phase transitions. As shown in Fig.~\ref{fig:bubble_profile}, regardless of the phase transition type and the value of $\omega$, as time progresses, the bubble radius increases and the wall thickness decreases. When the bubble walls are thin (top row of Fig.~\ref{fig:bubble_profile}), the scalar field near the bubble center remains close to the true vacuum in the quantum phase transition. While in the semiclassical phase transition, the scalar field exhibits very small oscillations around the true vacuum. This behavior arises from the fact that the exit point on the other side of the potential barrier after quantum tunneling/classical evolution is approaching the false vacuum in the thin-wall limit~\cite{Cai:2017tmh}. 
	
	After the bubble collision, the overlapping region of the two bubbles can rebound back and forth the potential barrier, similar to the trapping effect discussed in Ref.~\cite{Jinno:2019bxw}. When the bubble walls are thick (middle row of Fig.~\ref{fig:bubble_profile}), the trapping effect is weakened, but oscillations around the true vacuum become pronounced since the exit point is now on the potential hill away from the true vacuum, generating outgoing waves that follow the expanding bubble walls. As the wall thickness further increases (bottom row of Fig.~\ref{fig:bubble_profile}), a significant difference emerges in the semiclassical phase transition, where the scalar field evolution seems to resemble the expansion of a single bubble after collision. This phenomenon occurs as a large wall thickness corresponds to a small bubble separation (as shown in Table~\ref{tab:distance}), causing the classical perturbations to be so close that they can be nearly treated as a single perturbation. In this case, the production of gravitational waves is suppressed, as deviations from sphericity are small, but it is still intriguing for future study as a new way to generate the bubble, not from nucleation, but from the collision of false-vacuum classical perturbations. Nevertheless, in the following Sec.~\ref{sec:gravitational} and Sec.~\ref{sec:fitting}, we consider the semiclassical phase transition only for the case where two bubbles remain distinct (corresponding to $\omega < 0.9$).

	\section{Gravitational waves}\label{sec:gravitational}

	We now proceed to calculate the GW spectrum generated by the bubble collisions. Approaches to calculate the GW spectrum include the ``envelope approximation''~\cite{Kamionkowski:1993fg,Huber:2008hg,Jinno:2016vai} and the ``bulk flow model''~\cite{Konstandin:2017sat,Jinno:2017fby}. Here we closely follow the method adopted in Ref.~\cite{Lewicki:2019gmv} for a preliminary numerical calculations of the GW spectrum.

	\begin{figure*}
		\includegraphics[width=0.49\textwidth]{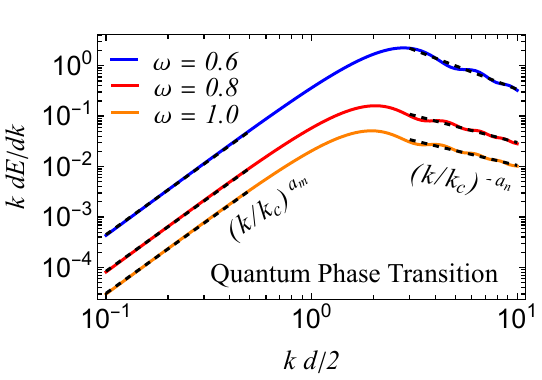}
		\includegraphics[width=0.49\textwidth]{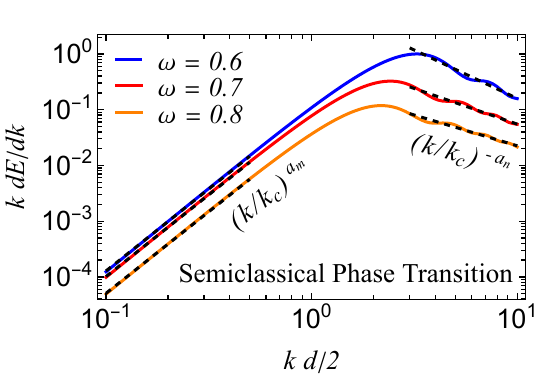}
		\caption{The gravitational wave spectrum $k \mathrm{d}E/\mathrm{d}k$ from two bubble collision as a function of GW frequency $k$. The left and right panels correspond to the quantum and semiclassical phase transitions, respectively. We normalize GW spectrums by setting the peak value of the spectrum in the semiclassical phase transition at $\omega = 0.6$ to 1.}
		\label{fig:GW}
	\end{figure*}

	The spectrum of GW is defined as~\cite{Weinberg:1972kfs,Jinno:2016vai,Lewicki:2019gmv,Cutting:2020nla}
	\begin{equation}
		\label{eq:gw_energy_0}
		\frac{\mathrm{d}E}{\mathrm{d} k} = 2Gk^2 \int \mathrm{d}\Omega\ \Lambda_{ijlm}(\hat{\mathbf{k}})T^{*}_{ij}(\hat{\mathbf{k}},k)T_{lm}(\hat{\mathbf{k}},k),
	\end{equation}
	where $G$ is the Newton constant, $\Omega$ is the solid angle, and $k$, equaling to the GW frequency, denotes the magnitude of the GW wave vector $\mathbf{k}\equiv k \hat{\mathbf{k}}$. Here, $\Lambda_{ijlm}$ is the projection tensor, 
	\begin{align}
		\label{eq:projection_tensor}
		\Lambda_{ijlm}(\hat{\mathbf{k}})=&\delta_{il}\delta_{jm}-2\delta_{il}\hat{\mathbf{k}}_j \hat{\mathbf{k}}_m+\frac{1}{2}\hat{\mathbf{k}}_i \hat{\mathbf{k}}_j \hat{\mathbf{k}}_l \hat{\mathbf{k}}_m \\ \nonumber
		&-\frac{1}{2}\delta_{ij}\delta_{lm}+\frac{1}{2}\delta_{ij}\hat{\mathbf{k}}_l \hat{\mathbf{k}}_m+\frac{1}{2}\delta_{lm}\hat{\mathbf{k}}_i \hat{\mathbf{k}}_j,
	\end{align}
	and the traceless part of the stress energy tensor generated by the scalar field is
	\begin{equation}
		\label{eq:stress_tensor}
		T_{ij}(\hat{\mathbf{k}},k)=\frac{1}{2\pi}\int \mathrm{d}t\mathrm{d}^3xe^{ik(t-\hat{\mathbf{k}} \cdot \mathbf{x})}\partial_i \phi \partial_j \phi.
	\end{equation}
	In this work, we consider the system of two identical bubbles, and the cylindrical symmetry in this system allows the integral over the azimuthal angle to be performed analytically. Furthermore, without loss of generality, we can set $\hat{\mathbf{k}} = (\sin \theta_k, 0, \cos \theta_k)$. The nonzero components of $T_{ij}$ are expressed as 
	\begin{widetext}
			\begin{align}
				\nonumber
				T_{\rho \rho}(\theta_k,k)&=-\int \mathrm{d}te^{ik t}\int_0^{\infty}\mathrm{d}z \cos\left(k \cos \theta_k z\right)\int_0^{\infty}\mathrm{d}\rho\ \rho\left[\sin^2 \theta_kJ_0\left(k \sin \theta_k \rho\right)+\left(\cos^2 \theta_k+1\right)J_2\left(k \sin \theta_k \rho\right)\right](\partial_{\rho} \phi)^2,\\
				\nonumber
				T_{\rho z}(\theta_k,k)&=-2\int\mathrm{d}te^{ik t}\int_0^{\infty}\mathrm{d}z\sin\left(k \cos\theta_k z\right)\int_0^{\infty}\mathrm{d}\rho\ \rho J_1\left(k \sin \theta_k \rho\right)\partial_{\rho} \phi \partial_z \phi,\\
				\label{eq:Tcomponent}
				T_{zz}(\theta_k,k)&=2\int \mathrm{d}te^{ik t}\int_0^{\infty}\mathrm{d}z\cos\left(k \cos \theta_k z\right)\int_0^{\infty}\mathrm{d}\rho\ \rho J_0\left(k \sin \theta_k \rho\right)(\partial_z \phi)^2,
			\end{align}
	\end{widetext}
	where $T_{\rho \rho}(\theta_k,k)\equiv \cos^2 \theta_k T_{xx}(\theta_k,k)-T_{yy}(\theta_k,k)$ and $J_i$ represent Bessel functions of the first kind. Now the integration in~\eqref{eq:gw_energy_0} reduces to
	\begin{widetext}
		\begin{equation}
			\label{eq:gw_energy_1}
			\frac{\mathrm{d}E}{\mathrm{d}k}=2\pi G k^2\int_0^{\pi}\mathrm{d}\theta_k \sin \theta_k |T_{\rho \rho}(\theta_k,k)+\sin^2 \theta_k T_{zz}(\theta_k,k)-2\sin \theta_k \cos \theta_k T_{\rho z}(\theta_k,k)|^2.
		\end{equation}
	\end{widetext}

	\begin{table*}[ht]
		\centering
		\renewcommand{\arraystretch}{1.25} 
		\setlength{\tabcolsep}{2.4pt} 
		\textbf{Quantum phase transition} \\[0.5em]
		\scalebox{1.2}{
			\begin{tabular}{*{12}{c}}
				\hline
				\hline
				$\omega$  & $k_c d/2$ & $E_{\rm p}$ & $A_m/A_n$ & $a_m$ &$a_n$ & $l_{\rm col}/R_{\rm col}$ & $\dot{l}_{\rm col}$ & $v_{\rm col}$ & $\sigma_{\rm col}$\\
				\hline
				0.60 & 2.85 & 2.25 & 3.48 &2.93 & 1.52 & 0.0086 & 0.061 & 0.993 & 0.00126 \\
				0.65 & 2.26 & 1.43 & 3.30 & 2.93 & 1.30 & 0.0150 & 0.060 & 0.987 & 0.00138\\
				0.70 & 2.15 & 0.54 & 3.23 & 2.93 & 1.21 & 0.0246 & 0.058 & 0.980 & 0.00157\\
				\hline
				0.75 & 2.06 & 0.33 & 3.65 & 2.92 & 1.11 & 0.0343 & 0.056 & 0.972 & 0.00163\\
				0.80 & 2.06 & 0.16 & 3.54 & 2.92 & 1.07 & 0.0478 & 0.055 & 0.964 & 0.00168\\
				0.85 & 1.96 & 0.12 & 3.43 & 2.92 & 1.02 &  0.0562 & 0.054 & 0.957 & 0.00167\\
				\hline
				0.90 & 1.96 & 0.07 & 3.20 & 2.92 & 1.00& 0.0645 & 0.053 & 0.950 & 0.00162\\
				0.95 & 1.96 & 0.06 & 3.52 & 2.92 & 0.98& 0.0736 & 0.051 & 0.945 & 0.00157\\
				1.00 & 1.96 & 0.05 & 3.53& 2.92 & 0.97 & 0.0804 & 0.050 & 0.940 & 0.00149\\
				\hline
				\hline
		\end{tabular}}
		\vspace{0.5cm}
		
		\textbf{Semiclassical phase transition} \\[0.5em]
		\scalebox{1.2}{
			\begin{tabular}{*{12}{c}}
				\hline
				\hline
				$\omega$ & $k_c d/2$ & $E_{\rm p}$ & $A_m/A_n$ & $a_m$ & $a_n$ & $l_{\rm col}/R_{\rm col}$ & $\dot{l}_{\rm col}$ & $v_{\rm col}$ & $\sigma_{\rm col}$\\
				\hline
				0.60 & 3.27 & 1.00 & 3.32 & 2.94 & 1.73& 0.0084 & 0.056 & 0.974 & 0.00125\\
				0.65 & 2.72 & 0.52 & 3.12 & 2.94 & 1.45 & 0.0151 & 0.053 & 0.968 & 0.00141\\
				0.70 & 2.36 & 0.32 & 3.12 & 2.93 & 1.28 & 0.0254 & 0.050 & 0.956 & 0.00162\\
				\hline
				0.75 & 2.26 & 0.18 & 3.21 & 2.93 & 1.19 & 0.0346 & 0.048 & 0.948 & 0.00164\\
				0.80 & 2.15 & 0.12 & 3.32 & 2.93 & 1.10 & 0.0455 & 0.045 & 0.941 & 0.00169\\
				0.85 & 1.96 & 0.11 & 3.23 & 2.92 & 1.01 & 0.0546 & 0.043 & 0.933 & 0.00164\\
				\hline
				\hline
		\end{tabular}}
		\caption{The numerical fitting results for the quantum phase transition and semiclassical phase transition. The first column is the potential parameter $\omega$. Parameters from the 2nd to the 6th columns are those shown in the parametrization~\eqref{eq:general_spectrum} of the GW spectrum. Here, we set $E_{\rm p}$ in the semiclassical phase transition with $\omega=0.6$ as 1 to normalize the peak value of GW spectra. The 7th column is the ratio of the bubble thickness to its radius at the collision time. The 8th to 10th columns are the wall thickness time derivative, the bubble wall velocity, and the bubble surface tension at the moment of bubble collision.}
		\label{tab:fitting}
	\end{table*}

	Using the scalar field profile $\phi(t,z,\rho)$ in Sec.~\ref{sec:collision}, we can compute $k \mathrm{d}E/\mathrm{d}k$ via~\eqref{eq:Tcomponent} and~\eqref{eq:gw_energy_1}, as shown in Fig.~\ref{fig:GW}. It is observed that there is no significant difference in the GW spectrum between the semiclassical and quantum phase transitions, both of which follow a simple broken power law. We model the spectrum as
	\begin{equation}
		\label{eq:general_spectrum}
		k\frac{\mathrm{d}E}{\mathrm{d}k}=E_{\rm p}\frac{(A_m+A_n)k^{a_m}k_c^{a_n}}{A_mk^{a_m+a_n}+A_nk_{c}^{a_m+a_n}},
	\end{equation}
	where $k_c$ denotes the peak frequency at which the spectrum reaches its maximum $E_{\rm p}$. Here, $A_m$, $A_n$, $a_m$, and $a_n$ characterize the power-law behavior of the GW spectrum. When $k \ll k_c$, the spectrum $\Omega_{\rm GW}$ follows the power law $k \mathrm{d}E/\mathrm{d}k\sim(k/k_c)^{a_m}$, whereas for $k \gg k_c$, it follows $k \mathrm{d}E/\mathrm{d}k\sim(k/k_c)^{-a_n}$. Numerical fitting results for the parameters in~\eqref{eq:general_spectrum} are presented in Table~\ref{tab:fitting}. It is easy to see that, at low frequencies, the $a_m$ values for different $\omega$ in both phase transitions are nearly identical, $a_m\approx3$, as expected from the causality requirement~\cite{Cai:2019cdl}. At high frequencies, however, the $a_n$ values vary depending on $\omega$ and the type of phase transition. Additionally, the peak values $E_{\rm p}$ of the spectrum for both phase transitions decreases as the bubble wall thickness increases, corresponding to higher $\omega$ values.

	\begin{figure*}
		\includegraphics[width=0.49\textwidth]{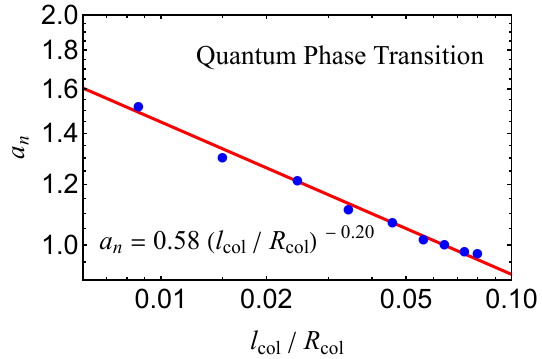}
		\includegraphics[width=0.49\textwidth]{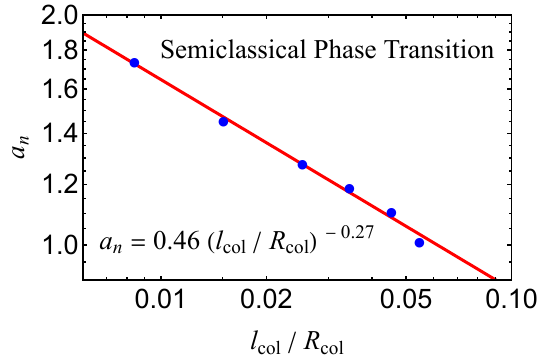}
		\caption{The numerical fitting results of the relationship between the high-frequency gravitational wave spectrum power and the ratio of bubble wall thickness to its radius at the collision time $l_{\rm col}/R_{\rm col}$. The blue dots are our simulation results and the red lines are plotted by the fitting function. The left and right panels correspond to the quantum and semiclassical phase transitions, respectively.}
		\label{fig:fitting}
	\end{figure*}

	\section{Numerical fitting}\label{sec:fitting}

	After obtaining GW spectra produced by the collision of two bubbles in Sec.~\ref{sec:gravitational}, we can now derive the numerical fitting relation between the high-frequency power law and $l_{\rm col}/R_{\rm col}$, the ratio of the bubble wall thickness $l_{\rm col}$ to the bubble radius $R_{\rm col}$ at the time of collision.

	To first determine the values of $l_{\rm col}$, we should obtain the bubble collision time. With the bubble profiles simulated in Sec.~\ref{sec:collision}, we can extract the time evolution of the bubble radius $R(t)$ and wall thickness $l(t)$. Then, compare $R(t)$ to the half of the bubble separation, $d/2 = R_{\rm col}$, the bubble collision time $t_{\rm col}$ is thus determined. Finally we obtain $l_{\rm col}\equiv l(t_{\rm col})$ and hence the ratio $l_{\rm col}/R_{\rm col}\equiv l(t_{\rm col})/R(t_{\rm col})$ at bubble collision. With all preparations completed, we numerically fit the relationship between the high-frequency power and $l_{\rm col}/R_{\rm col}$. The results are shown in Fig.~\ref{fig:fitting}. We observe that the absolute value of the power decreases as $l_{\rm col}/R_{\rm col}$ increases, which can be perfectly fitted with a single power law as
	\begin{equation}
		\label{eq:fitting_function}
		a_n=A_p\left(\frac{l_{\rm col}}{R_{\rm col}}\right)^{-a_p},
	\end{equation}
	\begin{align}
		\label{eq:quantum_result}
		(A_p,\ a_p)&=(0.5780,\ 0.1996),\ \ \ \mathrm{quantum\ PT},\\
		\label{eq:semiclassical_result}
		(A_p,\ a_p)&=(0.4647,\ 0.2747),\ \ \ \mathrm{semiclassical\ PT},
	\end{align}
	which should be only valid within the considered parameter range and may not be directly extrapolated into the thin-wall limit. This is because a naive extrapolation of this single power law into the thin-wall limit with $l_\mathrm{col}\to0$ would result in a divergent UV power $-a_n\to-\infty$, yielding a sharp but probably unphysical cutoff around the peak frequency.
	Recall that a previous study~\cite{Cutting:2020nla} has related $a_n$ intimately to the model potential parameter $\omega$. Here, we improve this study by fitting the high-frequency power to the more model-independent ratio, $l_{\rm col}/R_{\rm col}$, which will be further approximated in terms of other more model-independent characteristics of FOPTs in Sec.~\ref{sec:expression}. Nevertheless, it is worth noting that there is no theoretical guarantee that the high-frequency power should be related on its own to the ratio of bubble wall thickness and radius at collisions, and hence what we do here is purely phenomenological before fully understanding it analytically.
	
	Finally, it is important to note that Ref.~\cite{Gould:2021dpm} performs a similar numerical simulation for a two-bubble system, investigating the dependence of the high-frequency power $a_n$ on the scalar-field potential parameter and the bubble-wall Lorentz factor at collision time. The authors demonstrate that the difference in $a_n$ derived through various potential parameters becomes less pronounced at larger Lorentz factors. In contrast, in our study, once the scalar potential parameters are fixed, the mean bubble separation $d$ is also fixed, thereby determining the bubble Lorentz factor at collision time as a derived quantity rather than an independent parameter. Consequently, the explicit dependence of $a_n$ on the Lorentz factor is not apparent in our fitting result given by \eqref{eq:fitting_function}.

	\section{Analytical approximation}\label{sec:expression}
	
	In Sec.~\ref{sec:fitting}, we have derived the fitting results for the relationship between the high-frequency power and the wall thickness (normalized to the bubble radius) $l_{\rm col}/R_{\rm col}$ at collisions, as given in \eqref{eq:fitting_function}-\eqref{eq:semiclassical_result}. Although an immediate understanding of this result is currently not feasible, for specific FOPTs originating from different new physics models, calculating $l_{\rm col}/R_{\rm col}$ directly in terms of other more model-independent characteristics is still of large theoretical interest. Such calculations could help distinguish between various new physics models that lead to a first order phase transition.
	
	We estimate the value of $l_{\rm col}/R_{\rm col}$ from the energy conservation. The basic idea is that we first express the kinetic, gradient, and potential energies of a single bubble in terms of $l_{\rm col}$, $R_{\rm col}$, and other model-independent characteristics. Then by applying the energy conservation condition, we can obtain approximate expressions of $l_{\rm col}/R_{\rm col}$. Specifically, we first assume that before and at the moment of the bubble collision, the interaction between bubbles is negligible. Thus, the total energy is simply the sum of the energies of individual bubbles, each of which is conserved. Here, we make the following ansatz for the scalar profile $\phi$ of a single bubble,
	\begin{equation}
		\label{eq:scaler_ansatz_0}
		\phi=\frac{1}{2}\left[(\phi_-+\phi_+)-(\phi_--\phi_+)\tanh \frac{r-R(t)}{l(t)}\right],
	\end{equation}
	where $r$ represents the distance from the bubble center, $R(t)$ and $l(t)$ denote the bubble radius and wall thickness. Without loss of generality, we set $\phi_+=0$ and $\phi_-=\Delta \phi \equiv \phi_- - \phi_+$, so that the ansatz~\eqref{eq:scaler_ansatz_0} simplifies to
	\begin{equation}
		\label{eq:scaler_ansatz}
		\phi=\frac{\Delta \phi}{2}\left[1-\tanh \frac{r-R(t)}{l(t)}\right].
	\end{equation}
	Using~\eqref{eq:scaler_ansatz}, we obtain the field gradient as
	\begin{equation}
		\label{eq:gradient_scalar}
		\nabla \phi = -\frac{\Delta \phi}{2l(t)}\mathrm{sech}^2\left(\frac{r-R(t)}{l(t)}\right)\hat{r},
	\end{equation}
	where $\hat{r}$ is the unit vector in the $r$-direction. The gradient energy at collision~\eqref{eq:gradient_energy} is then approximated as
	\begin{align}
		\label{eq:gradient_energy_value}
		E_{G}(t_{\rm col})&=\frac{\pi}{2}\frac{(\Delta \phi)^2}{l_{\rm col}^2}\int_{0}^{\infty}\mathrm{d}r r^2 \mathrm{sech}^4\left(\frac{r-R_{\rm col}}{l_{\rm col}}\right)\nonumber\\ 
		&\approx \frac{2\pi}{3}(\Delta \phi)^2\frac{R_{\rm col}^2}{l_{\rm col}}.
	\end{align}
	Here, we have applied the approximation
	\begin{equation}
		\label{eq:integration_approximation}
		\int_{0}^{\infty}\mathrm{d}r r^2 \mathrm{sech}^4\left(\frac{r-R_{\rm col}}{l_{\rm col}}\right)\approx \frac{4R_{\rm col}^2 l_{\rm col}}{3},\ l_{\rm col}\ll R_{\rm col}.
	\end{equation}
	We then calculate $\partial_t \phi$ and approximate
	\begin{align}
		\label{eq:time_scalar}
		\partial_t \phi&=\frac{\Delta \phi}{2}\mathrm{sech}^2\left(\frac{r-R(t)}{l(t)}\right)\left[\frac{v_w l(t)+(r-R(t))\dot{l}(t)}{l^2(t)}\right]\nonumber\\ 
		&\approx \frac{\Delta \phi}{2l(t)}v_{w}(t)\mathrm{sech}^2\left(\frac{r-R(t)}{l(t)}\right),
	\end{align}
	where $v_w(t)\equiv \mathrm{d}R(t)/\mathrm{d}t$ is the bubble wall velocity. The reason for neglecting the second term in the square bracket is as follows: when evaluating the integral of $(\partial_t \phi)^2$ to obtain $E_{\rm K}$, the dominant contribution arises from the region around $r \sim R(t) \pm l(t)$. In this region, the factor $(r - R(t))$ in the second term is of order $\mathcal{O}(l(t))$. However, as explicitly demonstrated in Table~\ref{tab:fitting}, at the bubble collision time, the value of $\dot{l}(t)$ is smaller than $v_w$ by more than an order of magnitude. Consequently, the second term in the square brackets is significantly smaller compared to the first term, and thus can be safely neglected.
	\begin{align}
		\label{eq:kinetic_energy_value}
		E_{K}(t_{\rm col}) &= \frac{\pi}{2}\frac{(\Delta \phi)^2v_\mathrm{col}^2}{l_{\rm col}^2}\int_{0}^{\infty}\mathrm{d}r r^2 \mathrm{sech}^4\left(\frac{r-R_{\rm col}}{l_{\rm col}}\right) \\ \nonumber
		&\approx \frac{2\pi}{3}(\Delta \phi)^2v_\mathrm{col}^2\frac{R_{\rm col}^2}{l_{\rm col}}.
	\end{align}
	Note here that $v_\mathrm{col}=\sqrt{1-R_0^2/R_\mathrm{col}^2}$ has been evaluated as the bubble wall velocity at collision.
	We next express the potential energy~\eqref{eq:potential_energy} as
	\begin{align}
		\label{eq:potential_energy_value}
		&E_{V}(t_{\rm col}) =  \int_0^{\infty}\mathrm{d}r 4 \pi r^2 V(\phi(r, t_{\rm col})) \\  \notag
		&\simeq \int_0^{R_{\rm col}} \mathrm{d}r 4\pi r^2 V(\phi_-) + \int_{R_{\rm}+l_{\rm col}}^{\infty} \mathrm{d} r 4\pi r^2 V(\phi_+)  \\ \notag
		&+4 \pi R_{\rm col}^2 \int_{R_{\rm col}}^{R_{\rm col}+l_{\rm col}} \mathrm{d}r V(\phi(r, t_{\rm col}))  \\ \notag
		&\simeq -\frac{4\pi}{3}R_{\rm col}^3\Delta V+4\pi R_{\rm col}^2 \sigma^V_{\rm col}.
	\end{align}

Here, the second equality indicates that we have divided $E_{V}(t_{\rm col})$ into contributions from inside the bubble, outside the bubble, and the bubble surface region and have employed the approximation $l_{\rm col} \ll R_{\rm col}$. Furthermore, in deriving the third equality, we have used the conditions $V(\phi_+)=0$ and the definition $\Delta V = V(\phi_+)-V(\phi_-)$. In the last line of \eqref{eq:potential_energy_value}, we have separated out the potential contribution to the bubble surface tension at the collision time, $\sigma^V_{\rm col}$, explicitly given by
\begin{equation}
	\label{eq:sigma_define}
	\sigma^V_{\rm col}=\int_{R_{\rm col}}^{R_{\rm col}+l_{\rm col}} \mathrm{d}r V(\phi(r,t_{\rm col})).
\end{equation}
It is important to note that this contribution is part of the total contribution of the bubble surface tension~\cite{Linde:1981zj} that typically includes both potential and gradient energy contributions, i.e., $\sigma = \int \mathrm{d}r [(\mathrm{d} \phi / \mathrm{d} r)^2/2 + V(\phi)]$. 
The numerical values of $\sigma^V_{\rm col}$ obtained from our simulations are listed in the last column of Table~\ref{tab:fitting}. Our numerical results indicate that the second term in the final line of Eq.~\eqref{eq:potential_energy_value} is significantly smaller than the first term. Thus, in subsequent analyses, we approximate $E_V(t_{\rm col})$ simply as $-4\pi R_{\rm col}^3 \Delta V/3$.

	Now, the energy conservation $E_K+E_G+E_V=E_T$ can be expressed as
	\begin{equation}
		\label{eq:energy_conserve_new}
		\frac{2\pi}{3}(\Delta\phi)^2(1+v_\mathrm{col}^2)\frac{R_{\rm col}^2}{l_{\rm col}}-\frac{4\pi}{3}R_{\rm col}^3\Delta V=E_{T}.
	\end{equation}
	This gives
	\begin{equation}
		\label{eq:ratio_expression_one}
		\frac{l_{\rm col}}{R_{\rm col}}=\frac{\frac{2\pi}{3}(\Delta\phi)^2(1+v_\mathrm{col}^2)R_{\rm col}}{E_{T}+\frac{4\pi}{3}R_{\rm col}^3\Delta V}.
	\end{equation}
	We further simplify~\eqref{eq:ratio_expression_one} by neglecting $E_T$. This is because that, for the quantum phase transition, $E_T$ is simply zero, while for the semiclassical phase transition, the absolute value of the potential energy significantly exceeds the total energy as shown in Fig.~\ref{fig:energy} as the time evolves. Thus, we can safely approximate $E_{T}\ll 4\pi R^3_{\rm col}\Delta V$ and finally, Eq.~\eqref{eq:ratio_expression_one} reduces to
	\begin{equation}
		\label{eq:ratio_expression_two}
		\frac{l_{\rm col}}{R_{\rm col}}=\frac{(\Delta \phi)^2(1+v_\mathrm{col}^2)}{2R_{\rm col}^2\Delta V}.
	\end{equation}
	For our vacuum phase transition with runaway bubble, $v_\mathrm{col}=\sqrt{1-R_0^2/R_\mathrm{col}^2}$, and if the initial bubble radius $R_0$ at nucleation can be neglected compared to the bubble radius $R_\mathrm{col}$ at collision, this analytical approximation simply reduces to
	\begin{align}\label{eq:lbyRforvacuum}
		\frac{l_{\rm col}}{R_{\rm col}}&=\frac{(\Delta\phi)^2(2R_\mathrm{col}^2-R_0^2)}{2R_\mathrm{col}^4\Delta V}\approx\frac{(\Delta\phi)^2R_\mathrm{col}^{-2}}{\Delta V}\nonumber\\
		&=\frac{1+\alpha}{3\alpha}\left(\frac{1}{HR_\mathrm{col}}\right)^2\left(\frac{\Delta\phi}{M_\mathrm{Pl}}\right)^2,
	\end{align}
	where we have defined the phase hierarchy $\Delta\phi/M_\mathrm{Pl}\equiv\sqrt{8\pi G}\Delta\phi$ and strength factor $\alpha\equiv \Delta V/\rho_\mathrm{rad}$ along with the Friedmann equation $3H^2/8\pi G=(1+\alpha)\rho_\mathrm{rad}$. By combining the fitting results~\eqref{eq:fitting_function}-\eqref{eq:semiclassical_result} with~\eqref{eq:lbyRforvacuum}, we can easily break the parameter degeneracy from constraining the high-frequency power of the GW spectrum from vacuum bubble collisions.

	It is then tempting to guess the generalization of the above expression into the case of thermal phase transition with bubble wall velocity approaching a nearly constant terminal wall velocity. If most of these bubbles collide with each other before they would have reached the terminal wall velocity, the expression should be similar to the vacuum phase transition case~\eqref{eq:lbyRforvacuum} as most of the bubbles collide when they are still rapidly accelerating. If most of the bubbles collide long after they have approached the terminal wall velocity, then $v_\mathrm{col}$ should admit an extra dependence on the back-reaction force~\cite{Wang:2022txy,Wang:2023kux} (including the thermal-gradient force and fluid-friction force) acting on the wall. At the leading order, the backreaction pressure $\Delta p_\mathrm{LO}$ should be independent of the bubble velocity, and hence all the above derivation should be still valid as long as we replace $\Delta V$ with $\Delta V-\Delta p_\mathrm{LO}$ and treating $v_\mathrm{col}$ as an extra independent parameter in~\eqref{eq:ratio_expression_two}. In this case, one can define one extra parameter $\alpha_\infty=\Delta p_\mathrm{LO}/\rho_\mathrm{rad}$. Then, for an exponential nucleation rate $\Gamma\sim e^{\beta t}$, the mean bubble radius at collisions is given by $R_{\rm col}=2\pi^{1/3}v_\mathrm{col}/\beta$~\cite{Cutting:2020nla}, and the final expression~\eqref{eq:ratio_expression_two} becomes
	\begin{align}
		\label{eq:ratio_expression_three}
		\frac{l_{\rm col}}{R_{\rm col}}&=\frac{1+v_\mathrm{col}^2}{6(HR)_\mathrm{col}^2}\frac{1+\alpha}{\alpha-\alpha_\infty}\left(\frac{\Delta\phi}{m_\mathrm{Pl}}\right)^2\\
		&=\frac{1+v_\mathrm{col}^2}{24\pi^{2/3}v_\mathrm{col}^2}\frac{1+\alpha}{\alpha-\alpha_\infty}\left(\frac{\beta}{H}\right)^2\left(\frac{\Delta\phi}{M_\mathrm{Pl}}\right)^2.
	\end{align}
	
	Finally, we emphasize that our derivation relies on four key assumptions: (1) strict energy conservation for isolated bubbles before collision; (2) adopting the simplified scalar field profile given in \eqref{eq:scaler_ansatz_0}, valid mainly for relatively thin bubble walls; (3) bubble wall thickness being significantly smaller than bubble radius at collision; and (4) the total energy of each bubble being substantially smaller than the absolute potential energy magnitude. These assumptions collectively restrict our results to scenarios characterized by thin walls and large bubble separations. Therefore, our findings are limited and preliminary. In our future work, we will provide a more rigorous and generalized treatment of $l_{\rm col}/R_{\rm col}$ in terms of universally applicable model-independent characteristics.

	\section{Conclusions and discussions}\label{sec:conclusion}
	
	Investigating the stochastic GW signals offers a promising way to probe cosmological FOPTs and underlying new physics models. To gain a deeper understanding of the FOPT, it is crucial to analyze specific characteristics of the GW spectrum, such as high-frequency power law. In this work, we study the relationship between high-frequency power law and the ratio of bubble wall thickness to bubble radius at collision using numerical simulations and express this ratio as a function of model-independent parameters. Note that, to prepare the initial configurations for the numerical simulation, we derive for the first time the general expressions for the bubble number density and mean bubble separation in the vacuum first-order phase transition for runaway bubbles instead of a constant wall velocity. Additionally, to provide a more comprehensive analysis, we also consider the semiclassical phase transition and calculate for the first time the associated GWs.
	
	Although we have built the relationship~\eqref{eq:fitting_function} between the high-frequency power law and the bubble-wall thickness at collisions, the physical explanation for such a relation, along with the observed negative correlation, remains unclear. In the future, we aim to derive this relationship analytically, which will enhance our understanding of the dynamics of cosmological FOPTs. Furthermore, larger numerical simulations for collisions of more bubbles are also necessary to check our preliminary results to greater generality, though we expect the qualitative picture may remain the same. Finally, more general simulations should also be investigated for the thermal phase transition with the exponent in the decay rate beyond the linear order and with the given backreaction force beyond the leading order. In this case, the difficulty lies in the clean extraction from the total GWs for the pure scalar contribution of wall collisions, which is nevertheless subdominated to the sound waves.
	
	\begin{acknowledgments}
		This work is supported by 
		the National Key Research and Development Program of China Grants No. 2021YFC2203004, No. 2021YFA0718304, and No. 2020YFC2201501,
		the National Natural Science Foundation of China Grants No. 12422502, No. 12105344, No. 11821505, No. 12235019, No. 11991052, No. 12047503, No. 12073088, and No. 11947302,
		the Science Research Grants from the China Manned Space Project with No. CMS-CSST-2021-B01 and No. CMS-CSST-2025-A01. Our code and data have been made public on the website: https://github.com/Jun-Chen-Wang/Two-Bubble-Simulation/tree/main.
	\end{acknowledgments}

	\bibliography{ref}
	
\end{document}